# Quantitative CV-based indicators for research quality, validated by peer review

Nadine Rons[*] and Arlette De Bruyn[*]

[*]*Nadine.Rons@vub.ac.be, Arlette.De.Bruyn@vub.ac.be*
Research Coordination Unit, R&D Department, Vrije Universiteit Brussel, Brussels (Belgium)

**Introduction**

In a university, research assessments are organized at different policy levels (faculties, research council) in different contexts (funding, council membership, personnel evaluations). Each evaluation requires its own focus and methodology. To conduct a coherent research policy however, data on which different assessments are based should be well coordinated. A common set of core indicators for any type of research assessment can provide a supportive and objectivating tool for evaluations at different institutional levels and at the same time promote coherent decision-making. The same indicators can also form the basis for a 'light touch' monitoring instrument, signalling when and where a more thorough evaluation could be considered.

This poster paper shows how peer review results were used to validate a set of quantitative indicators for research quality for a first series of disciplines. The indicators correspond to categories in the university's standard CV-format. Per discipline, specific indicators are identified corresponding to their own publication and funding characteristics. Also more globally valid indicators are identified after normalization for discipline-characteristic performance levels. The method can be applied to any system where peer ratings and quantitative performance measures, both reliable and sufficiently detailed, can be combined for the same entities.

**Method**

From ex post peer review evaluations of research teams by international expert panels, two types of data sets are available: peer ratings on a series of aspects and quantitative performance measures for a series of standard output categories. Linear correlations between the two are calculated. Before doing so two types of normalizations are performed:
- Normalization for team size: The peer ratings are size-independent. To also create size-independent performance measures, these are calculated per full time equivalent leading staff (potential principal promoters of research projects).
- Normalization for discipline-characteristic performance levels: Both peer ratings and quantitative performance measures are normalized per discipline (aligning mean values and standard deviations), to correct for discipline-dependent evaluation and performance levels.

Performance measures that are significantly positively correlated with peer review results are selected as indicators for research quality. In order to avoid accidental occurrences, performance categories that are only present in the output of a minority of the teams are not taken into account.

**Material**

*Data & Research Disciplines*
Data were examined for a series of disciplines which were evaluated using the same standard methodology and which included a sufficient number of teams (involving evaluations finalized from 2000 to 2006).
Performance measures investigated in this context until now are different types of publications and project funding. These correspond to categories in the central research database of the university, also used to automatically extract researchers' CV's.
The quantitative peer review results include an overall evaluation as well as scores on scientific merit, planning, innovation, team quality, feasibility, productivity and scientific impact.

*Key Figures*
- 6 research disciplines and expert panels:
  Economics, Engineering, Informatics, Law, Philosophy & Letters, Political & Social Sciences
- 57 evaluated teams, 9 to 11 teams per discipline
- 263 full time equivalent postdoctoral level staff
- 63 experts from 11 countries
- 427 returned evaluation forms
- 8 peer review indicators
- 23 scientific publication categories from the university's CV-format + ISI-category
- 21 external project-funding categories

*Reliability*
To obtain significant correlations between results from different evaluation systems is not evident, as discussed by Moed (2005). Evaluations are designed to support particular decisions (e.g. funding) and do not necessarily consider aspects outside their focus, which however may be important in other evaluations. The peer review method used for this analysis produces peer ratings for a broad series of aspects and contains several precautionary measures to ensure reliable results (confidentiality, panel procedure, site visit, bias verification). It was designed in 1996-1997 taking into account recommendations and known problems following from earlier experiences as much as possible (Cozzens, 1997; Kostoff, 1997; Martin, 1996).

Reliability of the quantitative performance measures is ensured by data collection (for the files presented to the experts) in close collaboration between the central research administration and the teams.

**Findings & Discussion**
*Generally valid correlations*
Some publication categories are significantly positively correlated with peer ratings, globally as well as for almost all disciplines separately (without any significant negative correlation coefficients). This is observed for:
- "Articles in journals with international referee-system" and
- "Publications in journals indexed by the Thomson Scientific SCIE, SSCI or AHCI" (ISI-publications, largely overlapping with the previous category),
- Followed by "Communications at international conferences integrally published in proceedings".

This shows that even in domains where books are a prominent form of output, international, peer reviewed journal publications are a good indicator for research quality at team level.

Figure 1 shows how higher correlation coefficients are obtained after normalization per discipline (all disciplines included except Law for which different publication categories were used).

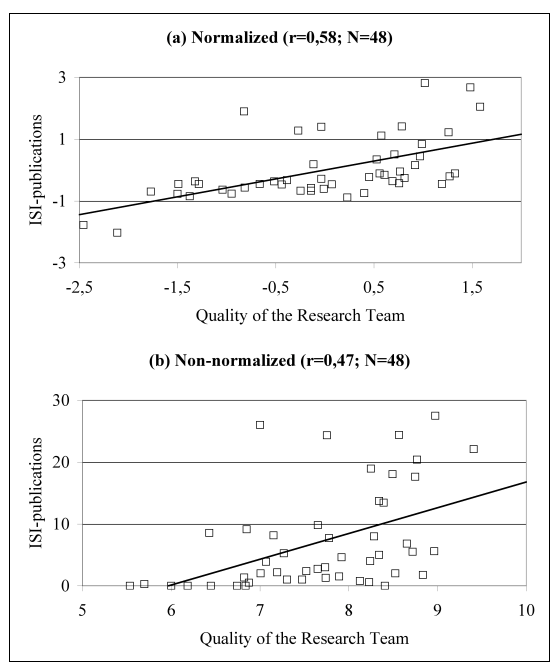

Figure 1: ISI-publications vs. peer ratings for "Quality of the Research Team".

*Differences between disciplines*
Disciplines differ in the particular categories of publications or project funding which are significantly correlated with peer ratings. These are in line with discipline-dependent typical funding channels (e.g. for applied or policy oriented research).

*Differences between performance categories*
Correlations with peer ratings differ for related performance categories, such as publications in journals with international, national or without referee system. It is therefore important to be able to distinguish between sufficiently fine categories in order to select appropriate indicators for evaluation. Broad performance categories may merge important performances with less important or even counterproductive ones. Obtaining significant correlations with such "mixed" performance measures is less evident and using them as indicators could be rewarding the wrong performances.

**Conclusions & Further Research**
This study for a first series of six disciplines shows that correlations between peer ratings and performance measures allow identifying core performance indicators, per research discipline as well as for larger research domains. Such a set of core indicators can be used as a common supportive tool for different kinds of evaluations, or it can be used in a monitoring instrument.

For evaluation purposes, the core performance indicators should be accompanied as much as possible by international reference values per discipline. International reference values however will not be available for locally defined performance categories. If also no national or regional reference values are available, averages within the institution could be constructed, provided a sufficiently large population is available.

Of course, while certain performance indicators may in general be related to quality as seen by peers, this does not necessarily imply these indicators' ability to distinguish between performances of individual researchers or even teams, unless correlations are perfect. Therefore, in the framework of an evaluation, interpretation of indicators by a committee remains necessary.

Future work will include an extension of the set of core performance indicators towards other disciplines (after results of their evaluations become available) and an investigation on reference values.